\documentclass[11pt,amsmath,superscriptaddress,nofootinbib]{revtex4}
\usepackage[english]{babel}

\usepackage[utf8]{inputenc}
\usepackage{floatrow}
\usepackage{amssymb,amsmath,amsfonts,amsthm,graphicx,psfrag}
\usepackage{indentfirst}
\usepackage{hyperref}

\usepackage[title,titletoc]{appendix}
\usepackage{bm}
\usepackage{verbatim}
\usepackage{epsfig}

\usepackage{mathrsfs}
\usepackage[T1]{fontenc}

\usepackage{slashed}
\usepackage{braket}
\usepackage{enumitem}

\newcommand{\GeV}{\text{GeV}}
\newcommand{\be}{\begin{eqnarray}}
\newcommand{\ee}{\end{eqnarray}}
\usepackage{color}

\newcommand{\figref}[1]{\figurename~\ref{#1}}

\begin{document}

\title{Quark-gluon plasma speed of sound in magnetic fields}

\author{Z.V.Khaidukov}
\affiliation{Moscow Institute of Physics and Technology, 9, Institutskii per., Dolgoprudny, Moscow Region, 141700, Russia}
\affiliation{Institute for Theoretical and Experimental Physics of NRC ``Kurchatov Institute'', B. Cheremushkinskaya 25, Moscow, 117259, Russia}
\author{R.A.Abramchuk}
\email{abramchuk@phystech.edu}
\affiliation{Moscow Institute of Physics and Technology, 9, Institutskii per., Dolgoprudny, Moscow Region, 141700, Russia}
\affiliation{Institute for Theoretical and Experimental Physics of NRC ``Kurchatov Institute'', B. Cheremushkinskaya 25, Moscow, 117259, Russia}

\begin{abstract}
    A non-perturbative method of Field Correlators (FCM) is applied to study speed of sound in quark-gluon plasma at temperatures above the deconfinement transition \(1<T/T_c<3\) under a uniform magnetic field up to \(eB\sim 1~\GeV^2\).
    The speed of sound in \(n_f=2+1\) QCD quark-gluon plasma is found to exceed the conformal limit ($c^{2}_{s}=\frac13$) at very large magnetic field strengths \(eB>0.5~\GeV^2\).
    At large temperatures, the speed of sound tends to the conformal limit;
    low-temperature behavior demonstrates inverse magnetic catalysis.
    The effect of magnetic field on the speed of sound in quark-gluon plasma is found to be negligible in any existing Heavy Ion Collision experiment, including \(\sqrt{s_{NN}}=5.02\) TeV Pb-Pb collisions at LHC.

%    %obj&scope
%   QGP in HIC at LHC energies/strong int nature
%   B~50m_pi^2 ~ 0.89 GeV^2 at LHC energies ~  TODO
%   strong B \to therm \to hydro \to observables ??
%   eB on centrality?
%    %methods
%   FCM + prev works quenched fermions + comp
%   ??isotropicity
%    %results
%   conformal limit exceeded at mod temp, respected at high temp
%    %conclusion
%   inverse magnetic catalysis
\end{abstract}
\maketitle

\section{Introduction}\label{sect_intro}

The incredible complexity of the Heavy Ion Collision (HIC) events is now addressed with a stack of (more or less effective) theories.
The fireball of quark-gluon plasma (QGP), which is a nearly ideal fluid, is described with relativistic hydrodynamics.
The hydrodynamic initial conditions and backgrounds are to incorporate details of a particular experimental setup, HIC or an astrophysical process.

The thermodynamic properties, equation of state and kinetic coefficients, are inputs for the hydrodynamic framework 
(note that equation of state remains relevant even for out-of-equilibrium fluid dynamics \cite{Romatschke:2017ejr};
i.e. in the Borel-resummed fluid dynamics, the non-equilibrium corrections are absorbed into shear and bulk stresses).
The properties may be obtained within various frameworks and models, e.g. Nambu-Jona-Lasono model, holographic models, linear response theory, lattice simulations. 
%pert QCD < thermodynamics of eff theory (non-pert QCD-inspired th) < hydrodynamics < hadronization model < decays + rescattering

Speed of sound is an essential parameter in a hydrodynamic model, yet a noteworthy QGP property.
It is well-defined at the thermodynamic limit, and was calculated with FCM \cite{Khaidukov:2018lor} in the absence of magnetic field.
On the other hand, it is immediately computed using HIC charged hadrons transverse momenta distribution and charged multiplicity data \cite{exp}.
%\begin{equation}
%    c_s^2 = \frac{dP}{d\epsilon}=\left.\frac{sdT}{Tds}\right|_{T_{eff}}=\frac{d\log\braket{p_t}}{d\log(dN_{ch}/d\eta)}.\label{c_s_exp}
%\end{equation}

An exceptional feature of HIC is emergence of extremely large magnetic fields (m.f.).
The averaged m.f.~influence on the hydrodynamics of QGP forming in HIC is imprinted in the elliptic flow \cite{bimprint}.
In the present paper, we study the influence on a deeper level --- we investigate if the m.f.~substantially affects thermodynamics of QGP.
In particular, if the m.f.~significantly affects speed of sound in QGP.

%We assume the parameter range to be relevant to high-energy Pb-Pb Heavy Ion Collisions at LHC.

The paper is organized as follows.
In section \ref{sect_sosb} we generalize the FCM speed of sound formula to the case of a nonzero m.f.
In section \ref{sect_num} we present a numerical evaluation of our analytical result.
Section \ref{sect_conc} contains conclusion and discussion.

\section{QGP speed of sound in magnetic fields}\label{sect_sosb}

The extremely strong initial m.f.~drops about few orders of magnitude during QGP formation process, which takes time \(t_0\sim1\) fm/c.
Before QGP formation, the initial m.f.~decays as in vacuum \(t^{-3}\), so the remnant m.f.~value is sensitive to QGP formation time \(t_0\).
However, \(t_0\) is an ill-defined quantity, so we are unsure of the magnetic field strength in a particular experiment.
We accept the conservative upper bound on m.f.~in a realistic experiment, including the \(\sqrt{s_{NN}}=5.02\) TeV Pb-Pb collisions at LHC, as \(eB<m_\pi^2\) \cite{indmf}.

The remnant m.f.~is 'frozen' in the QGP fireball due to its large electric conductivity.
Furthermore, the remnant m.f.~is dominant over the induced due to plasma motion, or valence, m.f. \cite{indmf}.
In what follows, we associate the external m.f.~with remnants of the initial m.f.

We start from thermodynamic description of QGP in magnetic field \cite{08}.
Assuming validity of isotropic picture, the pressure reads
\begin{align}
    P =& \frac{N_c e_q BT}{\pi^2}\sum_{q=u,d,s} \sum^\infty_{n=1} \frac{(-)^{n+1}}{n} L_f^n \left( M_{q} K_1 \left( \frac{nM_{q}}{T}\right)+\right.\\
    &\left.+\frac{2T}{n} \frac{e_qB+M_{q}^2}{e_qB} K_2 \left( \frac{n}{T} \sqrt{e_q B + M_{q}^2}\right) - \frac{ne_q
B}{12T} K_0\left( \frac{n}{T} \sqrt{M_{f}^2 + e_q B}\right)\right)+P_{gl} \label{teq},\\
    P_{gl}=&\frac{2(N^{2}_{c}-1)}{\sqrt{4\pi}}\int^\infty_0\frac{ds}{s^{3/2}}G_{3}(s)\sum^\infty_{n= 1}e^{-\frac{n^{2}}{4T^{2}s}}L^{n}_{adj},\\
    G_{3}(s)=&\frac{1}{(4\pi s)^{3/2}}\sqrt{\frac{M^{2}_{adj}s}{\sinh(M^{2}_{adj}s)}}  \label{basic}.
\end{align}
Here \(s\) is the proper time, \(B\) -- magnetic field,
\(N_{c}\) -- number of colors, \(e_{q}\) -- quark charge,
$M_{q}=\sqrt{m_{q}^2+{\frac14M_f^2}}$, \(q=u,~d,~s\) ,
$m_q$ -- bare quark mass.  
\(M_f^2=4a\sigma_{s}(T),~a\approx 1\) is the non-perturbative Debye mass, which emerges from color-magnetic confinement, and increases with temperature as 
$\sigma_{s}(T)=c_{\sigma}^{2}g(T)^{4}T^{2},\,c_{\sigma}=0.566\mp 0.013 $ \cite{Khaisim}.
%Debye mass of gluons is connected with this of a massless fermions via  .
%Polyakov loop  \(L\) is connected with interquark potential via :
%\be
%L=\exp\left(\frac{-V_{1}(\infty,T)}{2T}\right)
%\ee

QCD string tensions of strings in different gauge group representations are related according to the Casimir scaling law.
Thereby, the Polyakov loops and Debye masses in different representations are simply connected by the corresponding Casimir operators relation, in the present case --- adjoint (for gluons) and fundamental (for quarks):
$L_{adj}=L_f^{\frac94},\quad M^{2}_{adj}=\frac94M^{2}_{f}$.

We utilize the isotropic definition of speed of sound in QGP under magnetic field
\be
c^{2}_{s}=\left.\frac{dP}{d\epsilon}\right|_{B=\text{const}} \label{eq1},
\ee
where \(\epsilon\) is energy density
\be
\epsilon=T\frac{dP}{dT}-P .
\ee

%Another subtle question is if \(c_s\) is isotropic, since the external magnetic field discriminates a direction in space.%TODO

%For small fields, it is necessary to take into account the Debye modification of the quark masses.
%For fields with strengths over $ eB > 0.7~\GeV^{2}$ higher orders in \eqref{basic} become relevant, though the series is convergent due to the exponential suppression.
%}
%That happens mostly because the lowest Landau level dominates in this limit,
%For fields with strengths over $ eB > 0.7~\GeV^{2}$ higher orders in \eqref{basic} become relevant, though the series is convergent due to the exponential suppression. }

The major part of uncertainty is spawned by the neglected higher orders in the standard QCD perturbation theory, which is \(\sim\alpha_s^2\).
On top of the usual FCM approximations, we also neglect the Debye mass dependence on m.f.
Due to analytic structure of the formulae, the speed of sound is mildly affected by the relatively small Debye mass variance .
Moreover, the dependence is expected to be negligible even for large m.f.~\(eB\sim1\GeV^2\).
In result, we estimate the total error to be at most 15\% in the considered parameters range.

%\red{
%We neglect induced electromagnetic fields \cite{indmf}.???
%We also neglect the Debye mass dependencies on m.f.
%According to the recent lattice data, the approximations \(m_D(T,B)\to m_D(T)\) yield up to ?\% doesnt affect result due to exponent
%%\cite{09547}, 5\% \cite{00842}, and 2\% \cite{09461} errors, respectively.
%We consider the errors as independent, and estimate  the total \(c_s^2\) error as ?\%.
%Systematic error is $L(T, B)$ uncertainty 3-4%
%Within the parameters range, we estimate the total error of this section numerical results to be at most \(\%\).
%}

\section{Numerical results}\label{sect_num}

\begin{figure}%[h]

    \center{\includegraphics[width=0.7\linewidth]{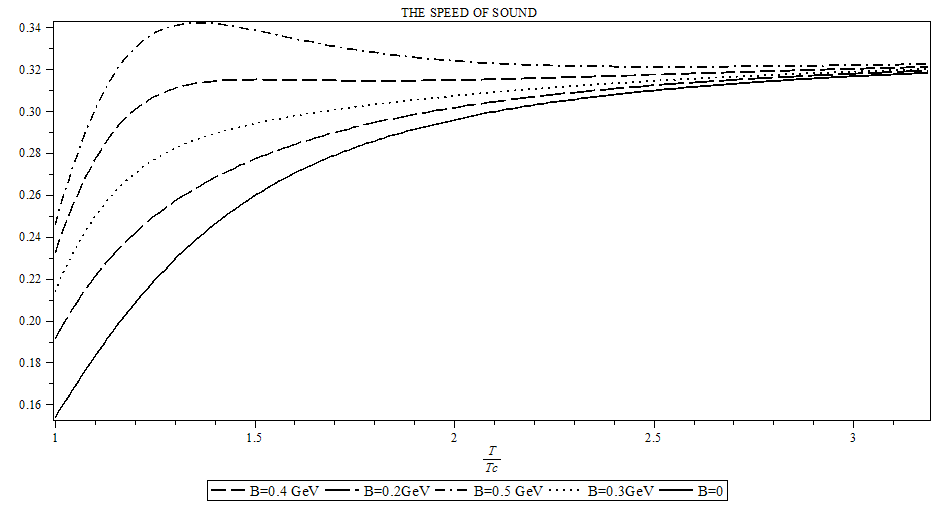}}
    \caption{Speed of sound squared in QGP as function of \(T/T_c\), where \(T_c=0.16~\GeV\) at various magnetic field strengths \(eB~[\GeV^2]\).}
    \label{fig_summ}
\end{figure}

\begin{figure}%[h]
    \center{\includegraphics[width=0.7\linewidth]{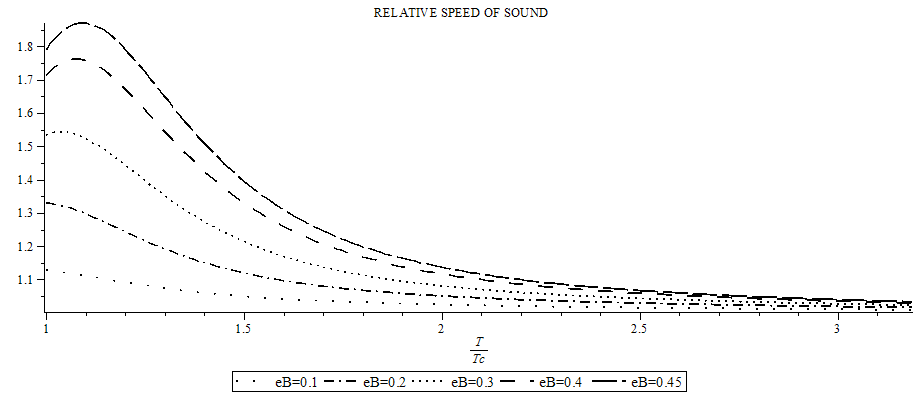}}
    \caption{Relative  square of the speed of sound as function of \(T/T_c\) at various \(eB~[\GeV^2]\).}
    \label{FiG2}
\end{figure}

\begin{figure}%[h]
\center{\includegraphics[width=0.7\linewidth]{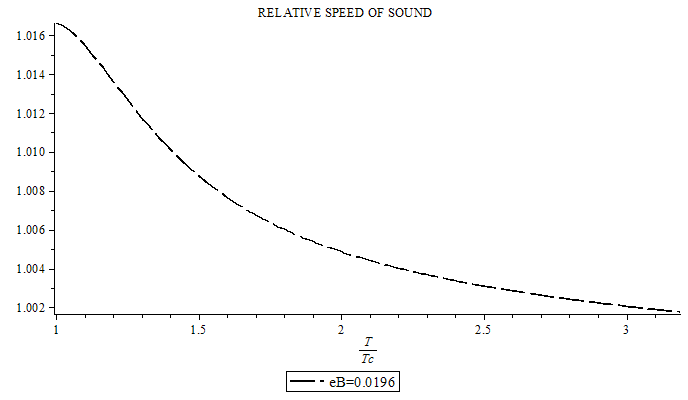}}
    \caption{Relative square of the  speed of sound  as function of \(T/T_c\) at $eB=m^{2}_{\pi}$.}
    \label{FiG3}
\end{figure}

We computed speed of sound in \(n_f=2+1\) QCD matter by means of the corresponding series \eqref{teq} partial summation.
We plot the speed of sound squared in \figref{fig_summ}, relation $\frac{c^{2}_{s,B}}{c^{2}_{s,0}}$ in \figref{FiG2} and \figref{FiG3}.

At large magnetic fields \(eB\sim1~\GeV^2\), we find the conformal limit excess to be reliably established.
However, at realistic magnetic fields \(eB\sim m_\pi^2\) the speed of sound change is overlapped by the uncertainties.

The Polyakov loop dependence \(L(T, B)\) was extracted from the lattice data \cite{PLB}.
The non-perturbative numerical input data were normalized \cite{08} as to reproduce the correct QGP pressure values at \(B=0\) in accordance with lattice simulations \cite{bor1,bor2,bor3,baz}.
The Polyakov loop uncertainty does not result in a significant uncertainty of the result.

\section{Conclusion and Discussion}\label{sect_conc}

We calculated the speed of sound in QGP under a uniform magnetic field.
We explicitly showed that for magnetic fields \(eB>0.4\GeV^2\) the speed of sound exceeds the conformal limit.
Though the Polyakov loop is sensitive to the magnetic field, the effect is dominated by the transverse quark motion quantization --- the effect is qualitatively reproduced even if the Polyakov loop dependence on magnetic field is neglected.
The low-temperature \(c_s^2\) dependence on \(B\) suggests inverse magnetic catalysis, which is in line with independent earlier predictions \cite{08,Braguta:2019yci};
however, the result is qualitative.

The principal result is mundane: the speed of sound changes significantly for large magnetic fields that are unattainable so far.
As for realistic  magnetic fields that emerge in LHC or RHIC collisions, the speed of sound variation is indistinguishable (see \figref{FiG3}) within our accuracy.
So hydrodynamic description with zero magnetic field speed of sound is perfectly applicable.

%Discussion
%Conclusion %   cl breaking at mod T %   cl respected at high T
%   inverse mag catalysis at low T
%   no diff for thermodynamics in today HIC

\section{Acknowledgments}
We are grateful to Yu.A. Simonov for useful discussions.
This work was supported by the Russian Science Foundation Grant number 16-12-10414.


\begin{thebibliography}{99}

%\cite{Romatschke:2017ejr}
\bibitem{Romatschke:2017ejr}
P.~Romatschke and U.~Romatschke,
        {\sl ``Relativistic Fluid Dynamics In and Out of Equilibrium,''}
doi:10.1017/9781108651998
[arXiv:1712.05815 [nucl-th]].
\bibitem{exp} Gardim, F.G., Giacalone, G., Luzum, M. et al.
%`` Thermodynamics of hot strong-interaction matter from ultrarelativistic nuclear collisions.''
    Nat. Phys. (2020). https://doi.org/10.1038/s41567-020-0846-4 [arXiv:1908.09728].
\bibitem{indmf} Kirill Tuchin, Phys. Rev. C 93, 014905 (2016) [arXiv:1508.06925].
\bibitem{bimprint} Long-Gang Pang, Gergely Endrődi, and Hannah Petersen, Phys. Rev. C 93, 044919 (2016) [arXiv:1602.06176].
%\bibitem{initb} Hattori, K., Huang, X. Novel quantum phenomena induced by strong magnetic fields in heavy-ion collisions. NUCL SCI TECH 28, 26 (2017). https://doi.org/10.1007/s41365-016-0178-3 , arXiv:1609.00747
%\cite{Khaidukov:2018lor}
\bibitem{Khaidukov:2018lor}
Z.~V.~Khaidukov, M.~S.~Lukashov and Y.~A.~Simonov,
%``Speed of sound in the QGP and an SU(3) Yang-Mills theory,''
Phys. Rev. D \textbf{98} (2018) no.7, 074031
doi:10.1103/PhysRevD.98.074031
[arXiv:1806.09407 [hep-ph]].
%5 citations counted in INSPIRE as of 25 Aug 2020
\bibitem{08} Abramchuk, R.A., Andreichikov, M.A., Khaidukov, Z.V., Simonov, Yu.A.,  Eur. Phys. J. C 79, 1040 (2019) [arXiv:1908.00800].
\bibitem{Khaisim}Z.V.Khaidukov,Yu.A.Simonov, Phys. Rev. {\bf D 100}, 076009 (2019)
    DOI: 10.1103/PhysRevD.100.076009 [arXiv:1906.08677]
\bibitem{bor1}
    S. Borsanyi, Z. Fodor, C.Hoelbling, S. D Katz, S. Krieg, C. Ratti,K.K. Szabo .,10.1007/JHEP {\bf  09}, 073 (2010), [arXiv:1005.3508].

\bibitem{bor2}{S.Borsanyi, Z.Fodor, C.Hoelbling,Phys. Lett. {\bf B 730},  99-104 (2014), [arXiv:1309.5258 [hep-lat]].}

\bibitem{baz}A.Bazavov,T.Bhattacharya, C. DeTar et al.,Phys. Rev.{\bf D 90}, 094503 (2014), [arXiv:1407.6387].
%25

\bibitem{bor3}S. Borsanyi, G.Endrodi, Z.Fodor, A.Jakovac, S. D. Katz, S.Krieg, C.Ratti, K.K. Szabo, JHEP {\bf 1011}, 077,(2010),
    [arXiv:1007.2580].
\bibitem{PLB}F. Bruckmann, G. Endrodi, T. G. Kovacs,JHEP 04 (2013) 112, [arXiv:1303.3972 [hep-lat]].

%\cite{Braguta:2019yci}
\bibitem{Braguta:2019yci}
V.~V.~Braguta, M.~N.~Chernodub, A.~Y.~Kotov, A.~V.~Molochkov and A.~A.~Nikolaev,
        {\sl``Finite-density QCD transition in a magnetic background field,''}
Phys. Rev. D \textbf{100} (2019) no.11, 114503
doi:10.1103/PhysRevD.100.114503
[arXiv:1909.09547 [hep-lat]].



\end{thebibliography}
\end{document}